\documentclass[%
 aip,
 jmp,%
 amsmath,amssymb,
 preprint,%
]{revtex4-2}

\usepackage{graphicx}
\usepackage{dcolumn}
\usepackage{bm}

\begin{document}


\title{Phase-change metasurfaces for reconfigurable image processing}

\author{Tingting Liu}
\affiliation{School of Information Engineering, Nanchang University, Nanchang 330031, China}
\affiliation{Institute for Advanced Study, Nanchang University, Nanchang 330031, China}

\author{Jumin Qiu}
\affiliation{School of Physics and Materials Science, Nanchang University, Nanchang 330031, China}

\author{Tianbao Yu}
\affiliation{School of Physics and Materials Science, Nanchang University, Nanchang 330031, China}

\author{Qiegen Liu}
\affiliation{School of Information Engineering, Nanchang University, Nanchang 330031, China}

\author{Jie Li}
\email{li_jie_d@tju.edu.cn}
\affiliation{Optoelectronic Sensor Devices and Systems Key Laboratory of Sichuan Provincial University, College of Optoelectronic Engineering, Chengdu University of Information Technology, Chengdu 610225, China}

\author{Shuyuan Xiao}%
\email{syxiao@ncu.edu.cn}
\affiliation{School of Information Engineering, Nanchang University, Nanchang 330031, China}
\affiliation{Institute for Advanced Study, Nanchang University, Nanchang 330031, China}

\date{\today}

\begin{abstract}
	Optical metasurfaces have enabled high-speed, low-power image processing within a compact footprint. However, reconfigurable imaging in such flat devices remains a critical challenge for fully harnessing their potential in practical applications. Here, we propose and demonstrate phase-change metasurfaces capable of dynamically switching between edge detection and bright-field imaging in the visible spectrum. This reconfigurability is achieved through engineering angular dispersion at electric and magnetic Mie-type resonances. The customized metasurface exhibits an angle-dependent transmittance profile in the amorphous state of Sb$_{2}$S$_{3}$ meta-atoms for efficient isotropic edge detection, and an angle-independent profile in the crystalline state for uniform bright-field imaging. The nanostructured Sb$_{2}$S$_{3}$-based reconfigurable image processing metasurfaces hold significant potential for applications in computer vision for autonomous driving systems. 
\end{abstract}

\maketitle


Optical analog computing has recently attracted considerable attention due to its ability to process images with high speed and low power consumption, particularly in real-time and high-throughput applications\cite{Solli2015,Wu2022}. All-optical image processing is a critical computational task that facilitates advancements in various applications such as augmented reality, virtual reality, and sophisticated surveillance systems. The traditional method to implement these operations is based on the complex filters and cumbersome 4f systems for the spatial filtering and transformation. These systems typically require precise alignment and suffer from significant limitations in terms of scalability and efficiency. The emerging artificially engineered nanostructures known as metasurfaces have been demonstrated to tailor electromagnetic properties within a compact footprint, enabling a variety of functionalities of focusing \cite{Wang2018,Yao2024,Li2024}, vortex \cite{Mei2023,Gou2024}, and hologram \cite{Yang2023,Yin2024,Liu2024}. Notably, metasurface-enabled image processing techniques, such as edge detection relying on differentiation operation in the spatial domain, have been widely explored \cite{He2022}. The presented approaches exploit various principles such as spin Hall effect \cite{Zhou2019,He2024}, surface plasmon polariton \cite{Zhu2017,Yang2020,Lou2020}, and Panchartnam-Berry phase \cite{Kim2021,Zong2023,Zhang2023}. In particular, nonlocal metasurfaces with angle-dependent responses have been studied theoretically and experimentally to perform momentum filtering of an image directly in real space \cite{Zhou2020,Pan2021,Komar2021,Ji2022,Cotrufo2023,Cotrufo2023a,Zhou2024,Liu2024b}, leading to the enhanced edges of input image while eliminating the need for a bulky 4f system, thus significantly reducing the system size. However, most of the differentiation metasurfaces are static with fixed functionalities that hinder their computing performance in more complex optical computing tasks. 

To overcome these limitations, various dynamic modulation strategies have been proposed to achieve integrated functionalities of image processing on the metasurface platform. One approach involves modulating input polarization on resonant metasurfaces for switchable optical differentiator \cite{Wang2023,Chen2023,Long2024}. For more versatile control of imaging functionality, some pioneering works have presented switchable analog computing by applying the external voltage on the graphene-based multilayer \cite{Momeni2022,Xia2023,Khodasevych2023} or the liquid crystal-based computing system \cite{Xiao2022,Liu2024a}. Then recent explorations incorporate active materials into metasurfaces, in order to realize tunable electromagnetic properties and thus reconfigurable image processing in a compact system. Phase-change materials (PCMs) have emerged as promising candidates for this purpose, exhibiting strong and controllable change in its refractive index induced by the phase transitions between the amorphous state and crystalline states \cite{Ding2019,Xiao2020}. PCMs such as VO$_{2}$, GST, and GSST have been investigated to integrate into optical computing metasurfaces for dynamic image processing \cite{Yang2021,Cotrufo2024,Heenkenda2021}. However, due to higher losses in the visible spectrum, the functionalities of these PCMs are predominantly limited to the infrared regime. Most recently, the emerging PCMs, including Sb$_{2}$S$_{3}$, Sb$_{2}$Se$_{3}$ and GeSe$_{3}$ with relatively low losses around 600 nm, have been successfully employed for high-resolution color and wavefront manipulation in the visible \cite{Dong2018,Delaney2020,Lu2021,Chen2021,Hemmatyar2021,Moitra2022,Liu2022}. Nevertheless, studies on reconfigurable image processing using such PCM-based metasurfaces in the visible remains scare.

In this work, we present Sb$_{2}$S$_{3}$-based metasurfaces capable of dynamically switching image processing functionalities between the edge-detection and bright-field imaging in the visible spectrum, with a tuning of the refractive index of the nanostructured Sb$_{2}$S$_{3}$ meta-atoms. The underlying principle of the reconfigurable metasurfaces lies in engineering the angular dispersion of electric and magnetic Mie-type resonances at different phase states of Sb$_{2}$S$_{3}$. The strong nonlocality of each of Mie-type resonances in the amorphous state implement a high-pass filter, enabling efficient, and isotropic edge detection. As Sb$_{2}$S$_{3}$ transforms into a fully crystalline state, angle-independent transmittance with reduced nonlocality of resonances leads to a nearly all-pass filter, facilitating uniform bright-field imaging. Especially, the proposed metasurfaces perform the dynamical image processing in a 4f-less configuration, significantly reducing the system footprint. The implementation of the reconfigurable image processing metasurfaces based on nanostructured Sb$_{2}$S$_{3}$ may find wide applications in computer vision for autonomous driving.


Figure 1 schematically illustrates the Sb$_{2}$S$_{3}$-based metasurfaces that can dynamically switch the image processing functionalities in the visible spectrum. When the input light field passes through the metasurface, 2D edge-detection and bright-field images can be obtained via adjusting phase states between the amorphous (a-) and crystalline states (c-) of Sb$_{2}$S$_{3}$. The two functionalities can be reversibly switched using appropriate thermal, electric, or optical stimulus \cite{Dong2018,Ding2019,Xiao2020}. The switchable imaging arises from tailoring the angular response of the metasurface for spatial filtering and transformation in Fourier space. Edge-detection and bright-field imaging functionalities necessitate distinct transfer functions (shown in the Supplementary Information, Section S1) from the angular responses for a- and c-Sb$_{2}$S$_{3}$, respectively. Specifically, the Laplacian operation for edge detection requires the optical transfer function of metasurface as $T(k_{||})=-k_{||}^2$, a quadratic function of the in-plane wavevector of the input light field. Conversely, to preserve the wavevectors and the original features of input images, the metasurface for bright-field imaging should have a transfer function keeping constant, i.e., $T(k_{||}) \propto c$, implying uniform transmission profile. Thanks to the large contrasts of the refractive index for a- and c-Sb$_{2}$S$_{3}$ metasurface, the angular responses in transmittance can be dramatically changed. In this context, we design a Sb$_{2}$S$_{3}$-based metasurface to satisfy the requirements of the transfer functions and the transmission responses for both functionalities.

\begin{figure*}[htbp]
	\centering
	\includegraphics
	[scale=0.45]{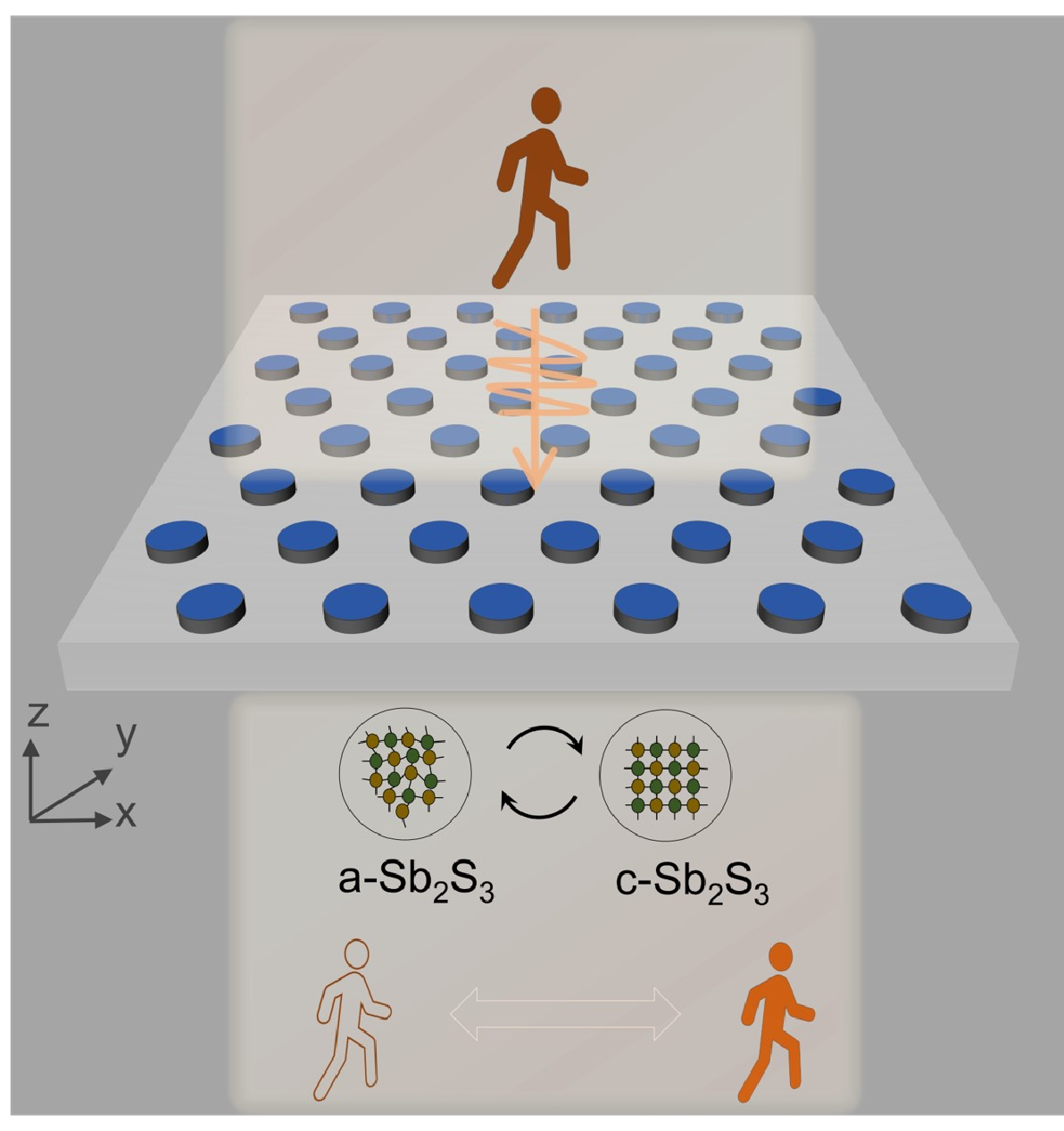}
	\caption{\label{Fig1} Schematic of the Sb$_{2}$S$_{3}$-based metasurface for reconfigurable imaging processing. The functionalities of the metasurface are dynamically switched between edge-detection and bright-field imaging in the visible spectrum via adjusting phase states of nanostructured Sb$_{2}$S$_{3}$ meta-atoms.}
\end{figure*}

The metasurface design is based on a hexagonal unit cell architecture, as shown in Fig. 2(a). It consists of Sb$_{2}$S$_{3}$ nanobricks of height $h$ and radius $r$ arranged in a hexagonal lattice with periodicity $a$ on a glass substrate. The design is numerically simulated using the finite-difference time-domain method to obtain the reconfigurable response at wavelengths in the visible spectrum (600 nm - 720 nm). The optical refractive index of Sb$_{2}$S$_{3}$ for both the amorphous and crystalline states are extracted from the experimental results \cite{Lu2021}, as shown in Fig. 2(b). It can be observed that within the wavelength of interest, the change of the real part of the refractive index $n$ between the two states can be as large as 0.8, and the imaginary part of refractive index $k$ equals 0 for a-Sb$_{2}$S$_{3}$ but show some optical losses with $k$ around 0.15 for c-Sb$_{2}$S$_{3}$. Accordingly, the nonlocal resonant property of the metasurface can be quite different, giving rise to the large contrast of the transfer function, thus bringing opportunities to implement different functionalities on a single metadevice. Then we optimize the geometrical parameters of the Sb$_{2}$S$_{3}$ metasurface such that it can work for edge detection with angle-sensitive transmittance in the amorphous phase and meanwhile satisfy bright field mode with an approximately flat transmittance profile in the crystalline state. 

With the optimized geometry, we calculate the transmission spectrum at normal incidence for both states, as shown in Fig. 2(c). In amorphous state, the spectra exhibit sharp resonance features, and the two separate transmission dips appear at 652 nm and 689 nm, respectively. The two resonances also determine operating wavelengths for image processing. At the two resonant wavelengths, the normal-incident transmission is smaller than 1$\%$, which offers ideal conditions to block the low wavevector components of the input image for enhanced edges. In contrast, the transmission spectrum of c-Sb$_{2}$S$_{3}$ metasurface is almost flat and remains around 0.3 within the wavelength regime. The much lower nonlocality is beneficial for the bright field mode in crystalline state. Owing to the $C_{6}$ rotational symmetry of the hexagonal design, the normal-incident transmission response is independent of the polarization states of illumination. To gain an insight of the physical origin of the strong resonances in a-Sb$_{2}$S$_{3}$ metasurface, we perform the multipolar decomposition with the results shown in Fig. 2(d). According to the decomposition results, the magnetic dipole (MD) mode plays dominant role in the resonance around 652 nm, and at the same time, the electric dipole (ED) mode provides primary contributions to the resonance at 689 nm. These results of the main multipolar contributors prove the Mie-type nature of the two resonances. 

\begin{figure}[htbp]
	\centering
	\includegraphics
	[scale=0.5]{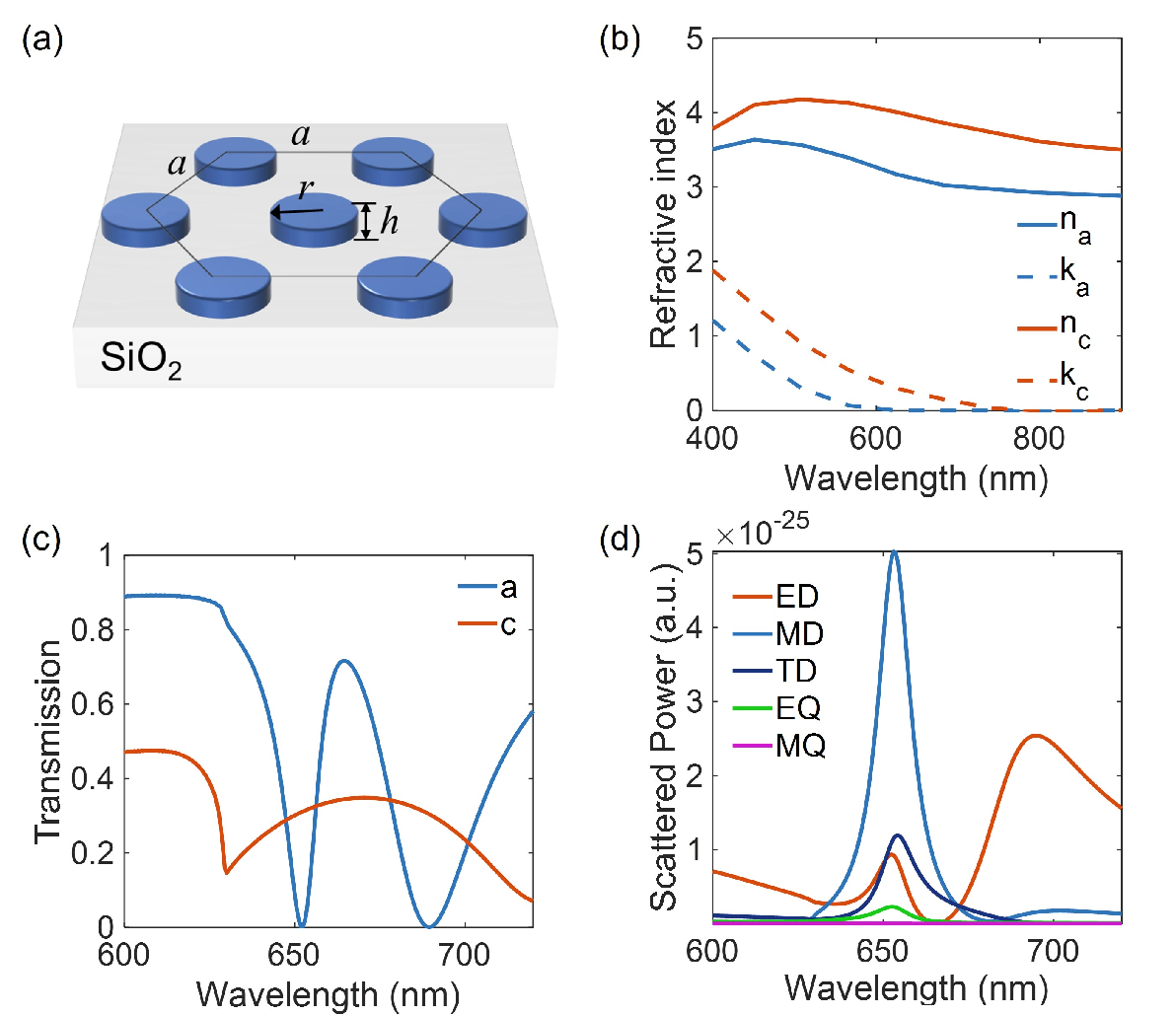}
	\caption{\label{Fig2} (a) Schematic of a unit cell of the metasurface with optimized geometry of $a$=500 nm, $h$=100 nm, and $r$=160 nm.(b) Refractive index of Sb$_{2}$S$_{3}$ in the amorphous and crystalline states. (c) Simulated transmission spectra under normal incidence for both states. (d) Results of multipolar decomposition for amorphous state.}
\end{figure}

For the two states of Sb$_{2}$S$_{3}$, we investigate the angular dispersion at the respective wavelength of electric and magnetic resonances. We numerically calculate the transmission amplitude ($|t|$, the square root of transmission) at the resonant wavelengths of 652 nm and 689 nm as a function of incident angle, as shown in Figs. 3(a) and 3(d), respectively. At both wavelengths, a-Sb$_{2}$S$_{3}$ metasurface show strong  dependence of transmission amplitude on incident angle, arising from the lowest electric and magnetic dipole modes of Mie-type resonances. It can also be found that the transmission amplitude satisfy the required quadratic dependence on the incident wavevector ($k\propto \text{sin}\theta$), which is a fit data with the function of $|t|_\text{fit}= A \text{sin}\theta ^{2} $. The maximum fit of the quadratic curve can reach $\theta \sim 10^{\circ}$ with NA $\sim$0.17 for the electric resonance, indicating an edge detection resolution around 2.4 $\mu$m. The transmission amplitude can be as high as $\sim$87$\%$ at the oblique incident angle of 10$^{\circ}$ at the electric resonance, suggesting high differentiation efficiency. In the same angular range, the transmission amplitude for c-Sb$_{2}$S$_{3}$ metasurface remains stable, with the values changing from 0.5$\sim$0.6. The similar case applies to the magnetic resonance as well. Such angle-independent response indicates that the metasurface would preserve the spatial wavevectors despite some amplitude reduction in the crystalline state. 

We also present the transmission amplitude at the two operating wavelengths as function of the wavevectors at different azimuthal angles. Without loss of generality, we focus on the results under $x$-polarization states of incident light here. As illustrated by Figs. 3(b) and 3(e), a-Sb$_{2}$S$_{3}$ metasurface at both wavelengths features high-pass filter, ensuring that the low wavevector components of the incident light can be completely suppressed and the high wavevectors can be transmitted in any directions. The nearly circular profile of the 2D transmission spectra with insensitivity to the arbitrary azimuthal angle from 0 to 360$^{\circ}$ also suggests the reproduction of almost the desired transfer function $|t|\propto k_{||}^{2}=k_{x}^{2}+k_{y}^{2}$, implying the differentiation along both directions, i.e., nearly isotropic edge detection. When the Sb$_{2}$S$_{3}$ is in crystalline state, the transmission amplitude at both wavelengths keeps almost a constant value with invariance to different wavevectors, as shown in Figs. 3(c) and 3(f). The metasurface behaves as an all-pass filter, which faithfully keep all the wavevectors of the input light. Therefore, the Sb$_{2}$S$_{3}$ metasurface satisfies the requirements of the transfer functions and the transmission responses, so that it can work for edge detection in the amorphous state and for bright-field imaging in the crystalline state. 

\begin{figure}[htbp]
	\centering
	\includegraphics
	[scale=0.6]{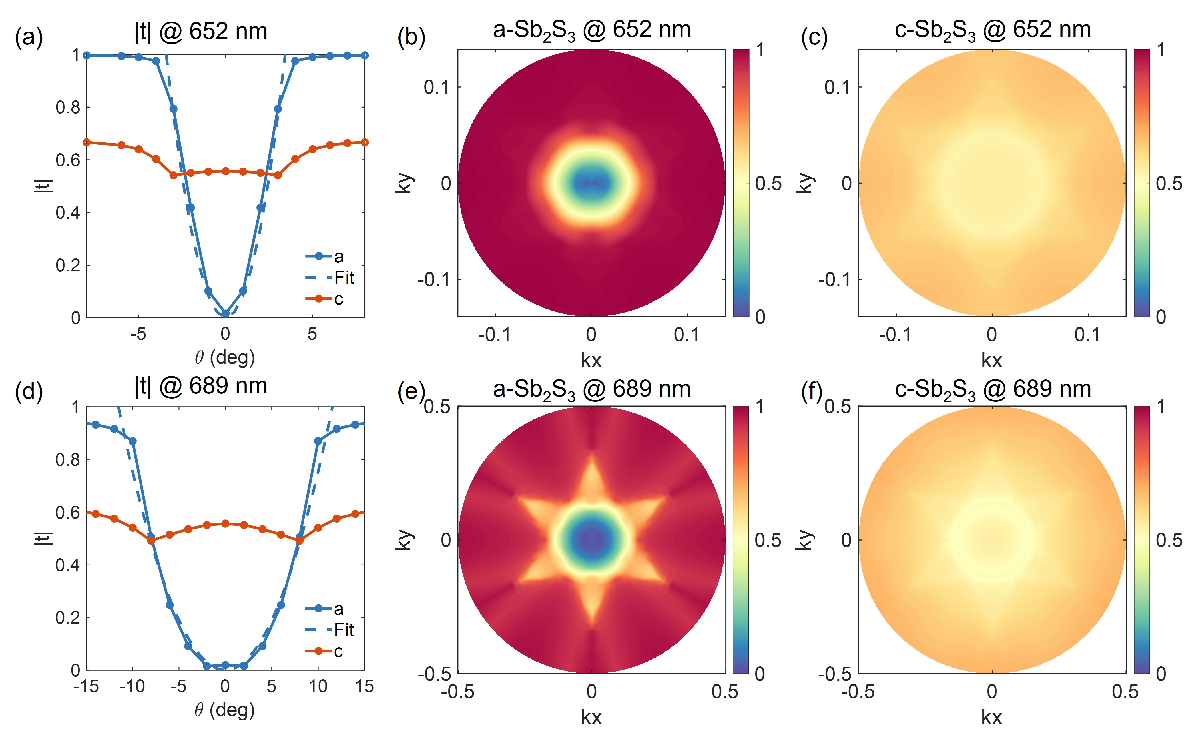}
	\caption{\label{Fig3}  Angular dispersion of the metasurface at resonant wavelengths of 652 nm and 689 nm for amorphous and crystalline states under $x$-polarization incidence. (a)(d) Transmission amplitude as a function of incident angle. (b)(c)(e)(f) The results of optical transfer functions.}
\end{figure}

To verify the reconfigurable image processing functionality of Sb$_{2}$S$_{3}$ metasurface, we firstly investigate the dynamical imaging functionality on 2D images from a resolution test chart. The images consist of three vertical and horizontal stripes as shown in Figs. 4(a) and 4(e), respectively. Since the designed metasurface can be functioned as a specific spatial filter in different phase states, the input images are modulated by the metasurface and the output results can be directly obtained by simply recording the transmitted light field instantaneously. On this basis, we calculate the Fourier spectrum of the input image $E_\text{in}(x,y)$, and then apply the transfer function $T(k_{x},k_{y})$ in Fig. 3 to the Fourier spectrum, yielding an output light filed $E_\text{out}(x,y)$ through an inverse Fourier transformation at the image plane. This image processing is expressed mathematically as $E_\text{out}(x,y)= \text{IFT}\{T(k_{x},k_{y})\text{FT}[E_\text{in}(x,y)]\}$. With this processing procedure, we calculate the output field profiles of the proposed metasurface in both amorphous and crystalline states. For a-Sb$_{2}$S$_{3}$ metasurface, the edges of the stripes are clearly revealed along the horizontal and vertical directions, indicating high-quality and isotropic 2D spatial differentiation. As Sb$_{2}$S$_{3}$ metasurface phase changes to the crystalline state, output results faithfully reproduce the input images due to the all-pass filter response. And the lower transmission amplitude also results in some intensity attenuation after passing the metasurface compared with the input images. The intensity profiles across the middle of the input and output images are presented in Figs. 4(d) and 4(h). The high-intensity peaks at the expected edge positions are observed in amorphous state, further verifying the high-performance edge detection of the a-Sb$_{2}$S$_{3}$ metasurface. In contrast, there is no significant difference in the profiles with and without c-Sb$_{2}$S$_{3}$ metasurface, and the good agreement suggests the fidelity in bright-field imaging. It is also noted that the dynamical switching between the edge-detection and bright-field imaging is realized at the wavelength of 652 nm and 689 nm, corresponding to the magnetic and electric Mie-type resonances, respectively.

\begin{figure}[htbp]
	\centering
	\includegraphics
	[scale=0.6]{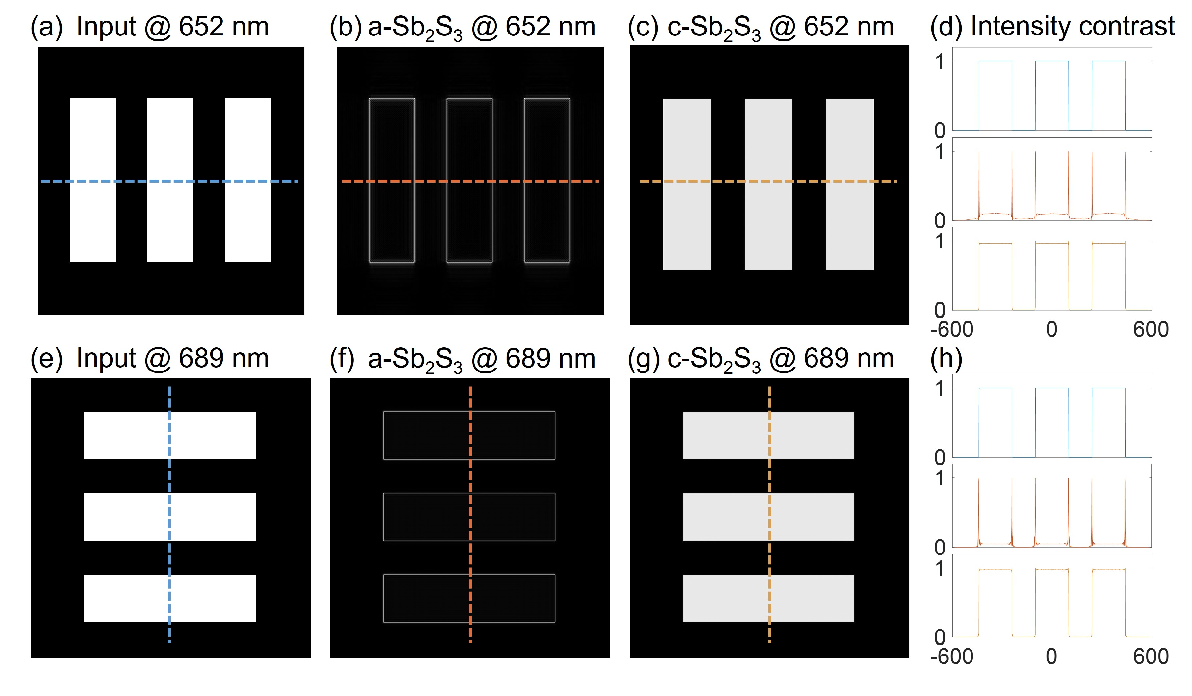}
	\caption{\label{Fig4} Imaging results of a test chart and the corresponding horizontal/vertical intensity profiles at resonant wavelengths of 652 nm and 689 nm for amorphous and crystalline states. }
\end{figure}

\begin{figure}[htbp]
	\centering
	\includegraphics
	[scale=0.5]{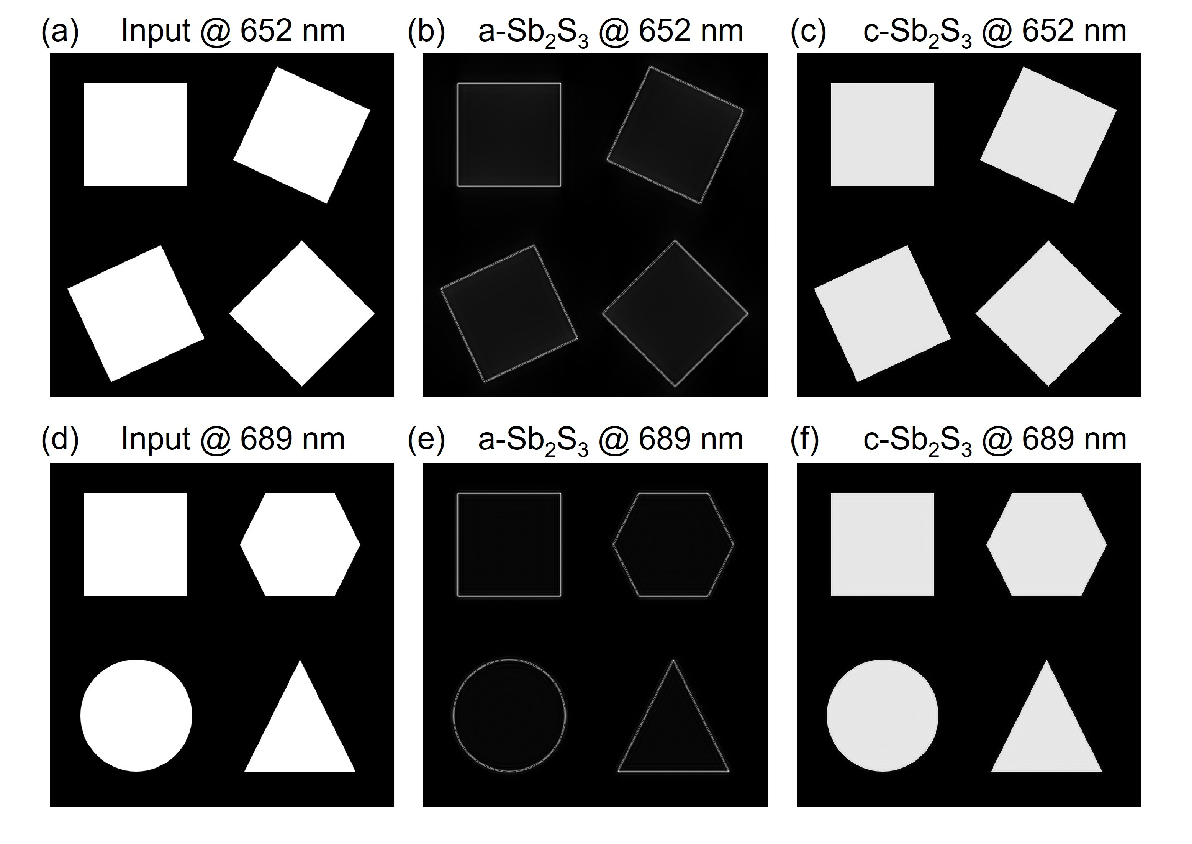}
	\caption{\label{Fig5} Imaging results of different shapes at resonant wavelengths of 652 nm and 689 nm for amorphous and crystalline states.  }
\end{figure}

Then we compute imaging results of a mask with different shapes to further validate the high-performance and isotropic image processing functionalities, as shown in Fig. 5. After passing through a-Sb$_{2}$S$_{3}$ metasurface, the input image is automatically processed, and the filtered image displays a clear, isotropic, and high-contrast edge enhancement. According to intensity distributions in Fig. S1 (shown in the Supplementary Information, Section S2), the edge detection effect of image shows high quality with the intensity of the detected edges nearly 10 times higher than that of the background. The results for c-Sb$_{2}$S$_{3}$ metasurface which reproduces the input images confirm the efficient bright-field imaging functionality. Besides the excellent reconfigurability, the image processing metasurface also exhibit good isotropic operations in either functionality. For amorphous state, all the edges in different directions are equally enhanced, consistent with the ideal second-order Laplace operation. The similar effect can also be observed in bright-field imaging for c-Sb$_{2}$S$_{3}$. Therefore, the reconfigurable metasurface achieves ideal Laplacian filtering and all-pass filtering, leading to direction-independent, isotropic and uniform image processing.

\begin{figure}[htbp]
	\centering
	\includegraphics
	[scale=0.6]{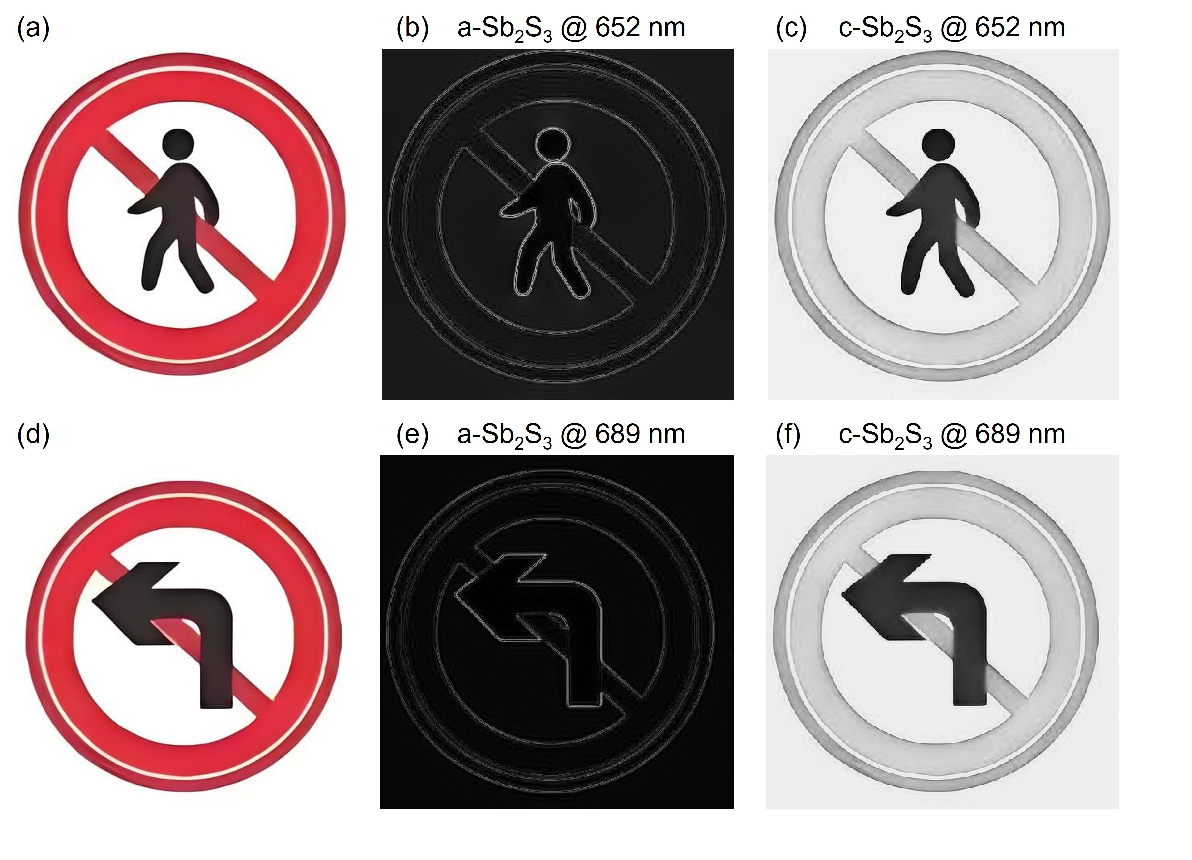}
	\caption{\label{Fig6} Input and output images of two common traffic signs at resonant wavelengths of 652 nm and 689 nm for amorphous and crystalline states. }
\end{figure}

In the realm of image processing, one of the most prevalent applications is found in the field of autonomous driving, which involves the real-time analysis of images to identify pedestrians, vehicles, and obstacles in the surrounding environment. Here we demonstrate the great potential of the proposed metasurfaces in recognizing different traffic signs, where high-performance image processing can be achieved without the need of a 4f system. In practice, such image processing configurations can be considered as an optical processor, which can not only faithfully record the unfiltered input image using the bright-field imaging functionality, but also can perform analog edge detection of the image in high performance using the Laplacian filtering. We select some general traffic signs as test objects and provide their original version in Figs. 6(a) and 6(d). When these images with modulated intensity are impinged on the proposed metasurface, a high-efficient and high-contrast edge enhancement can be observed in amorphous state, while the signs are recorded in their original images in crystalline state. We also notice that the edges of the circle, arrow and pedestrian are clearly identified, significantly improving object recognition accuracy and image clarity. During this process, the important information is securely retained while irrelevant details are disregarded, aids in reducing the complexity of image data and making subsequent analysis and recognition tasks more efficient.


In conclusion, we have proposed a novel approach to realize reconfigurable image processing by tailoring the nonlocality of the proposed Sb$_{2}$S$_{3}$-based metasurface. In our design, the angle sensitivity of transmission profile can be dramatically controlled by tuning the refractive index of the nanostructured Sb$_{2}$S$_{3}$ meta-atoms, enabling switchable functionalities between edge-detection and bright-field imaging. The reconfigurability of the metasurface stems from engineering the angular dispersion of electric and magnetic Mie-type resonances at different phase states of Sb$_{2}$S$_{3}$. In amorphous state of Sb$_{2}$S$_{3}$, the metasurface features an angle-dependent transfer function that facilitates isotropic edge detection. When Sb$_{2}$S$_{3}$ transitions to its fully crystallinity, an angle-independent profile with reduced nonlocality is obtained for a uniform bright-field imaging. The proposed imaging metasurface manifests the advantage of high efficiency, reduced complexity, and the switchable ability to implement two different transfer functions. Importantly, the proposed metasurface does not require the use of a 4f lens system, since the operations are implemented by filtering wavevectors directly in real space. The design opens up new opportunities for multifunctional and miniaturized processor, with promising applications in biological imaging and computer vision. 

\section*{Data availability}

Relevant data supporting the key findings of this study are available within the article and the Supporting Information file. All raw data generated during the current study are available from the corresponding authors upon reasonable request.

\section*{Conflicts of interest}

The authors declare no conflict of interest.

\begin{acknowledgments}
	
This work was supported by the National Natural Science Foundation of China (Grants No. 12304420, No. 12264028, No. 12364045, and No. 12404484), the Natural Science Foundation of Jiangxi Province (Grants No. 20232BAB201040 and No. 20232BAB211025), and the Young Elite Scientists Sponsorship Program by JXAST (Grants No. 2023QT11 and 2025QT04).

\end{acknowledgments}



\nocite{*}

%

\end{document}